\documentclass[final,3p,times,onecolumn]{elsarticle}

\usepackage[utf8]{inputenc}
\usepackage{textcomp}
\usepackage{lmodern,textcomp} % euro sign0
\usepackage{subfig}
\usepackage[-]{callouts}
\usepackage{tikz}
\usetikzlibrary{decorations.pathreplacing,calligraphy}

\graphicspath{{graphics/}}

\DeclareGraphicsExtensions{.pdf,.jpeg,.png}

\usepackage{epsfig}
\usepackage{lipsum}

\usepackage{amsmath}
\usepackage{amsfonts}
\usepackage{amssymb}
\usepackage{float}
\usepackage[hyphens]{url}

\usepackage[normalem]{ulem}

\usepackage{booktabs}
\usepackage{tabularx}
\usepackage{threeparttable}
\usepackage{siunitx}
\usepackage[version=4]{mhchem}
\usepackage[normalem]{ulem}

\usepackage{url}
\usepackage[colorlinks=true, citecolor=blue, linkcolor=blue, filecolor=blue,urlcolor=blue,breaklinks=true]{hyperref}
\usepackage{enumitem}

\usepackage[gen]{eurosym}
\usepackage{multirow}

\usepackage{threeparttable}
\usepackage{makecell}

\usepackage{color, colortbl}
\makeatletter
\newcommand{\vast}{\bBigg@{3}}
\newcommand{\Vast}{\bBigg@{4}}
\makeatother

\usepackage{tikz}
\usepackage{adjustbox}

\usepackage{fixltx2e}
\usepackage{pdflscape}
\usepackage{changepage}

\begin{document}
%\linenumbers
\begin{frontmatter}

\title{Accounting for carbon capture solvent cost and energy demand in the energy system}

\author[chalmers]{V.~Chanal\footnotemark[1]\corref{cor1}}
\ead{vincentc@chalmers.se}
\author[chalmers]{S.~Humpage\footnotemark[1]}
\author[chalmers,rise]{M.~Millinger}

\cortext[cor1]{Corresponding author}
\address[chalmers]{Department of Space Earth and Environment, Chalmers University of Technology, Göteborg, Sweden}
\address[rise]{Built Environment: System Transition: Energy Systems Analysis, RISE Research Institutes of Sweden, Göteborg, Sweden}
\footnotetext[1]{These authors contributed equally}

\begin{abstract}
% MTh abstract
Technical carbon dioxide removal through bioenergy with carbon capture or direct air capture plays a role in virtually all climate mitigation scenarios. Both of these technologies rely on the use of chemical solvents or sorbents in order to capture CO$_2$. Lately, concerns have surfaced about the cost and energy implications of producing solvents and sorbents at scale. Here, we show that the production of chemical sorbents could have significant implications on system cost, energy use and material use depending on how much they are consumed. Among the three chemical sorbents investigated, namely monoethanolamine (MEA) for post-combustion carbon capture, potassium hydroxide for liquid direct air capture and polyethylenimine-silica (PEI) for solid sorbent direct air capture, we found that the production of the compound for solid sorbent direct air capture represent the highest uncertainties for the system. At the high range of solid sorbent consumption, total energy system cost increased by up to 6.5\%, while effects for other options were small to negligible. Scale-up of material production capacities was also substantial for MEA and PEI. Implications of sorbent consumption for carbon capture technologies should be considered more thoroughly in scenarios relying on direct air capture using a solid sorbent.
\end{abstract}

\begin{keyword}
direct air capture \sep post-combustion \sep CCU \sep CCS \sep sorbent
\end{keyword}

\end{frontmatter}

\section{Introduction}
Carbon capture plays a key role in virtually all climate mitigation scenarios limiting the global mean temperature increase to 1.5$^o$C and 2$^o$C compared to pre-industrial levels \cite{babiker2022cross}. Captured carbon can be sequestered to avoid fossil emissions to the atmosphere, or to provide carbon dioxide removal (CDR) if the carbon stems from biomass (BECCS) or direct air capture (DACCS). Alternatively, utilisation of captured carbon (CCU) provides raw material to production of fuels and chemicals, and contributes to increasing carbon efficiency of scarce biogenic carbon \cite{Millinger}. In the IPCC Illustrative Mitigation Pathways that limit warming to 2$^o$C or lower between 168–763~GtCO$_2$ and 0–339~GtCO$_2$ are sequestered in this century through BECCS and DACCS, respectively \cite{babiker2022cross}, corresponding to an annual average of 6-26\% and 0-11\% of 2024 global CO$_2$ emissions \cite{ieaco2em2023}. Studies with sector-coupled energy system models also include CCU and arrive at 1-6 Gt annual carbon capture by mid-century globally \cite{GALIMOVA2022133920, IEA2020CCUS}, or 390-800 Mt for Europe \cite{PICKERING20221253,LUT,hofmann2024h2co2networkstrategies}.

However, carbon capture technology has yet to be proven at scale. 51 Mt carbon capture capacity is in place in 2024, mainly in fossil applications for enhanced oil recovery \cite{globalccsinstitute2024status}. 30 Mt bioenergy with carbon capture (BECC) \cite{IEAbeccs} and 65 Mt DAC \cite{IEAdaccs} is being planned world-wide. Besides expansion challenges \cite{Kazlou2024}, high cost and energy demand and resource limitations \cite{Millinger}, carbon capture requires the use of solvents or sorbents which capture CO$_2$ from either syngas (pre-combustion), flue gases (post-combustion), or from the atmosphere (direct air capture, DAC).

While solvents for post-combustion carbon capture have been proven at commercial scale, solvents and sorbents for DAC have been less researched and deployed. Uncertainties regarding costs \cite{young2023cost,SIEVERT2024979} as well as energy and material requirements \cite{Unrealistic} related to solvents and sorbents are substantial. For example, Chatterjee \& Huang \cite{Unrealistic} estimated that the energy demand of material production and solvent regeneration alone may amount to 12-20\% and 34-51\% of global energy demand, respectively, signaling that these aspects deserve a closer analysis.

Despite the large role played by carbon capture in modeled mitigation scenarios, uncertainty regarding solvents has thus far received little attention, and costs and energy use of solvent production has not explicitly been taken into account in such studies. This study aims to quantify the effect of cost and energy use uncertainties of solvents and sorbents for carbon capture on the cost-competitiveness of carbon capture technologies and on energy system cost.

\section{Methods}

\subsection{Data}
\begin{figure}[H]
    \centering
    \includegraphics[width=\linewidth, height= 0.5 \linewidth]{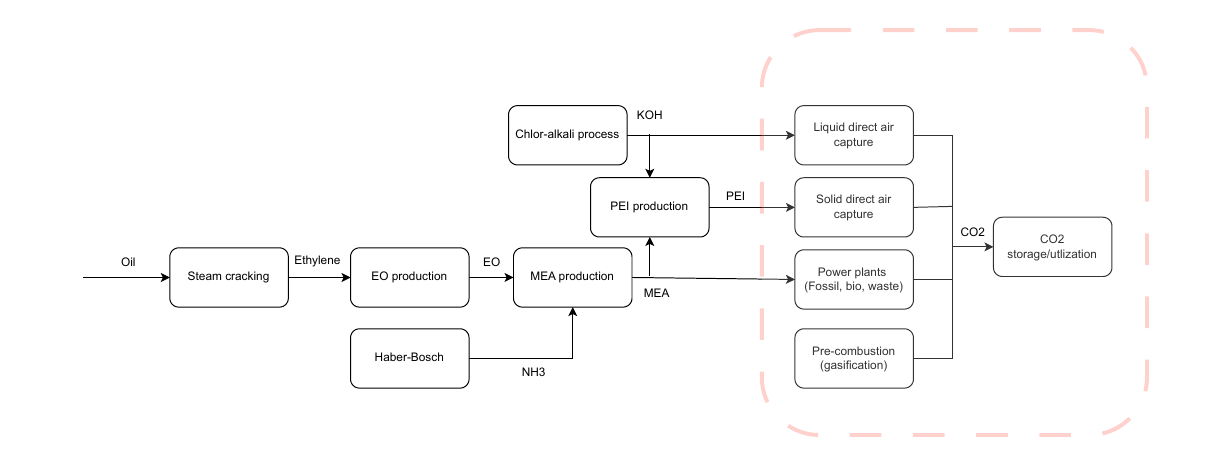}
    \caption{Model representation of solvent and sorbent production in this study. The base version with less detail is highlighted by red dashed box.}% Post-combustion as well as pre-combustion capture, which take place in power or chemical plants, only represent the carbon capture infrastructure of the plant, without the plant itself, that is modelled separately.}
    \label{fig:Production_flowchart}
\end{figure}

This study focuses on three different carbon capture technologies each using one solvent or sorbent: post-combustion carbon capture using a monoethanolamine (MEA) solvent; L-DAC, a high-temperature direct air capture using the liquid solvent potassium hydroxide (KOH); S-DAC, a low-temperature direct air capture using a polyethylenimine (PEI) on silica solid sorbent. The solvent and sorbent production methods were estimated based on currently available technologies (Fig. \ref{fig:Production_flowchart}).

S-DAC and L-DAC data (except those related to solvent and sorbent consumption) account for technological learning and economies of scale for a total scale of 1 Gt of global carbon capture capacity, based on Young et al. \cite{young2023cost}. Post-combustion carbon capture data were obtained from the Danish Energy Agency database \cite{DEA_tech_catalogues}. The provision and consumption of solvents or sorbents are isolated from the other costs and considered separately in the analysis.

% \paragraph{Solvent consumption}
In this work, an optimistic and a pessimistic scenarios represent the lower and the higher bounds for costs and energy use. Post-combustion MEA consumption ranges between 0.27-3.98~kg/tCO$_2$ \cite{DUVALDACHARY2023113415}. Consumption values of KOH for L-DAC ranges between 0.4~kg/tCO$_2$ \cite{NASEM,Socolow} and 38~kg/tCO$_2$ \cite{madhu2021understanding} and are assumed to represent the value range. S-DAC sorbent consumption ranges between 2.3-14~kg/t$_{CO_2}$, based on an estimated lifetime range of 0.5-3~years \cite{madhu2021understanding}. Another estimate based on data from Climeworks of 3-7.5~kg/tCO$_{2}$ \cite{deutz} falls within this range, although the full range of 2.3-14~kg/tCO$_{2}$ is used in this work to represent the uncertainty. Data for chemicals production cost and feedstock is detailed in the \nameref{suppinfo}.

\subsection{Model}
PyPSA-Eur-Sec \cite{Brown2023,Victoria2022} is an open-source, sector-coupled European energy system optimisation model including the power sector, transport (including also international shipping and aviation), space and water heating, industry and industrial feedstocks. The model minimises total system costs by co-optimising capacity expansion and operation of all energy generation and conversion, as well as of storage and transmission of electricity, hydrogen and gas. The model is based on the Python software toolbox PyPSA (Python for Power Systems Analysis) \cite{Brown2018c}. A comprehensive description of the model can be found in Neumann et.al. \cite{Neumann2023a}. A version with an extended biomass and technology portfolio is used \cite{Millinger, Millinger2022b}.

A 37-node spatial resolution and a 5-hourly temporal resolution over a full year in overnight green-field scenarios was used. A lossy transport model for electricity transmission was used which is suitable at this resolution \cite{Neumann2022}, and transmission is constrained to expand to at most double the total line volume in 2022.

Final energy demands for the different sectors are calculated based on the JRC IDEES database \cite{Mantzos2017} with additions for non-EU countries \cite[see][for further elaboration]{Victoria2022, Zeyen2021}, and need to be met (i.e. demand is perfectly inelastic). However, energy carrier production including electricity, hydrogen, methane and liquid fuels is determined endogenously. Fossil fuels (coal, natural gas and oil) as well as uranium are included, as are solid biomass imports as outlined below. Technology costs and efficiencies are elaborated on in the supplementary information, with technology values for 2040 (given in \texteuro$_{2015}$) used from the PyPSA energy system technology data set v0.6.0 \cite{TechnologyData2023}. The discount rate is uniform across countries, and set to 7\%, except for rooftop solar PV and decentral space/water heating technologies, for which it is set to 4\%.

\subsection{Scenarios}
Based on their effect on carbon capture and large variety of assumptions in climate mitigation scenarios, three factors are varied, resulting in 8 main scenario combinations: fossil fuel usage, biomass availability and emission target.

Climate mitigation scenarios differ considerably in how much fossil fuels are used, which has a strong impact on the utilisation of CCS and CCU for achieving targets. Some scenarios rely heavily on CCS which allows substantial amounts of fossil fuels to be used, while others assess 100\% renewable energy systems in which case more CCU is used. In order to represent both of these approaches, scenarios both including and excluding fossil fuels are explored. 

Previous assessments also differ strongly in the assumed biomass availability. Therefore, scenarios with only domestic residues from the JRC ENSPRESO medium scenario \cite{JRC2021} and ones including also biomass imports are explored. Biomass imports are assumed to represent a carbon neutral resource with a substantially higher price \cite{Millinger}, which could however also be supplied by extending the use of domestic resources.

The European Union has aimed to become the first continent to reach net zero emissions across all sectors and to do so by 2050 \cite{EuropeanUnion2021}. Net-negative targets in the energy system may be needed to offset hard-to-abate emissions from, for instance, agriculture, as well as to compensate for historic emissions \cite{IEA2020CCUS}. Therefore, net-zero and -110\% net-negative emissions targets are analysed.

In order to better be able to compare with previous model assessments which thus far mostly do not include L-DAC, these 8 scenarios are explored in cases where L-DAC is included and excluded, with both pessimistic and optimistic solvent and sorbent assumptions.

\subsection{Cost assessment}
\label{cost_assessment}

As a method of comparing the effects that implementation of solvent and sorbent production has on the cost of carbon capture systems, their Levelized Cost of Carbon (LCOC) with and without accounting for solvents are compared. This metric represents the cost of capturing one unit of CO$_2$. For direct air capture plants, the LCOC is calculated according to Eq.~\ref{eq:LCOC_DAC}.

\begin{equation}
LCOC_{DAC}=\frac{CAPEX \cdot (CRF+F_{OM})}{8760 \cdot C_f}+V_{OM}+\delta_{el} \cdot p_{el}+\delta_H \cdot p_{th}
    \label{eq:LCOC_DAC}
\end{equation}

CAPEX is the capital cost of the investment in €/MW$_{output}$. It is annualised using a capital recovery factor, the calculation of which is given by Equation \ref{eq:CRF}. Then, F$_{OM}$ is the fixed operations and maintenance cost [\% of CAPEX /year], 8760 the number of hours per year, C$_f$ is the capacity factor, V$_{OM}$ is the variable operations and maintenance cost in [€/MWh$_{output}$], which excludes cost of energy and solvent makeup, $\delta_{el}$ is the electricity demand [MWh], p$_{el}$ is the electricity price [€/MWh], $\delta_H$ is the heat demand [MWh], p$_{th}$ is the heat price [\texteuro/MWh]. In the impact assessment, solvent production cost is included in the V$_{OM}$.
\begin{equation}
    CRF=\frac{i(1+i)^n}{(1+i)^n-1} 
    \label{eq:CRF}
\end{equation}
where i represents the discount rate [\%/year] and n the lifetime of the plant [years].

The LCOC for power plants (LCOC$_{pp}$) is calculated as the LCOE difference with LCOE$_{CC}$ and without LCOE$_{NoCC}$ [\texteuro/MWh$_{output}$] carbon capture divided by carbon captured $\eta_{cc}$ [t$_{CO_{2}}$/MWh$_{output}$] \cite{GARCIA2022108470} (Eq.~\ref{eq:LCOC}).

\begin{equation}
LCOC_{pp}=\frac{LCOE_{CC}-LCOE_{NoCC}}{\eta_{cc}} 
\label{eq:LCOC}
\end{equation}

The LCOE of a power plant measures the cost of producing one unit of energy \cite{GARCIA2022108470} (Eq.~\ref{eq:LCOE}).

\begin{equation}
     LCOE=\frac{CAPEX \cdot (CRF+F_{OM})}{8760 \cdot C_f}+V_{OM}+\frac{C_{fuel}}{\eta}
    \label{eq:LCOE}
\end{equation}

C$_{fuel}$ is the cost for fuel in €/MW$_{th}$ and $\eta$ the efficiency of the plant. For power plants producing both heat and electricity, the LCOE is calculated per unit of total energy output. Therefore, $\eta = \eta_{el} + \eta_{th}$.

% \subsection{Energy assessment}
% To assess the implications of solvent production on the technologies, the energy requirements for said production is compared to the energy requirements for operating the CC facility, i.e. to regenerate the solvent or sorbent, since it is the main source of energy consumption for CC technologies \cite{GARCIA2022108470}. If one neglects any other energy requirements in the plant's lifetime (e.g., during the plant construction, or dismantlement), the sum of both these consumptions correspond to the CC facility’s life-cycle energy requirement.

% For the impact assessment on the entire system, a comparison is made between the energy required for producing all the chemicals implemented to the total primary energy consumed in the system.

% =================================================================================

\section{Results and discussion}
The system effect of accounting for solvent costs and energy use is threefold: first, the specific energy input differs for the various solvents; second, solvent replenishment rates are subject to large uncertainties, leading to large variances in costs and energy use; third, the deployment of different carbon capture options affects the cost contribution of solvents and sorbents.

\begin{figure}[H]
    \centering
    \includegraphics[width=0.7\linewidth]{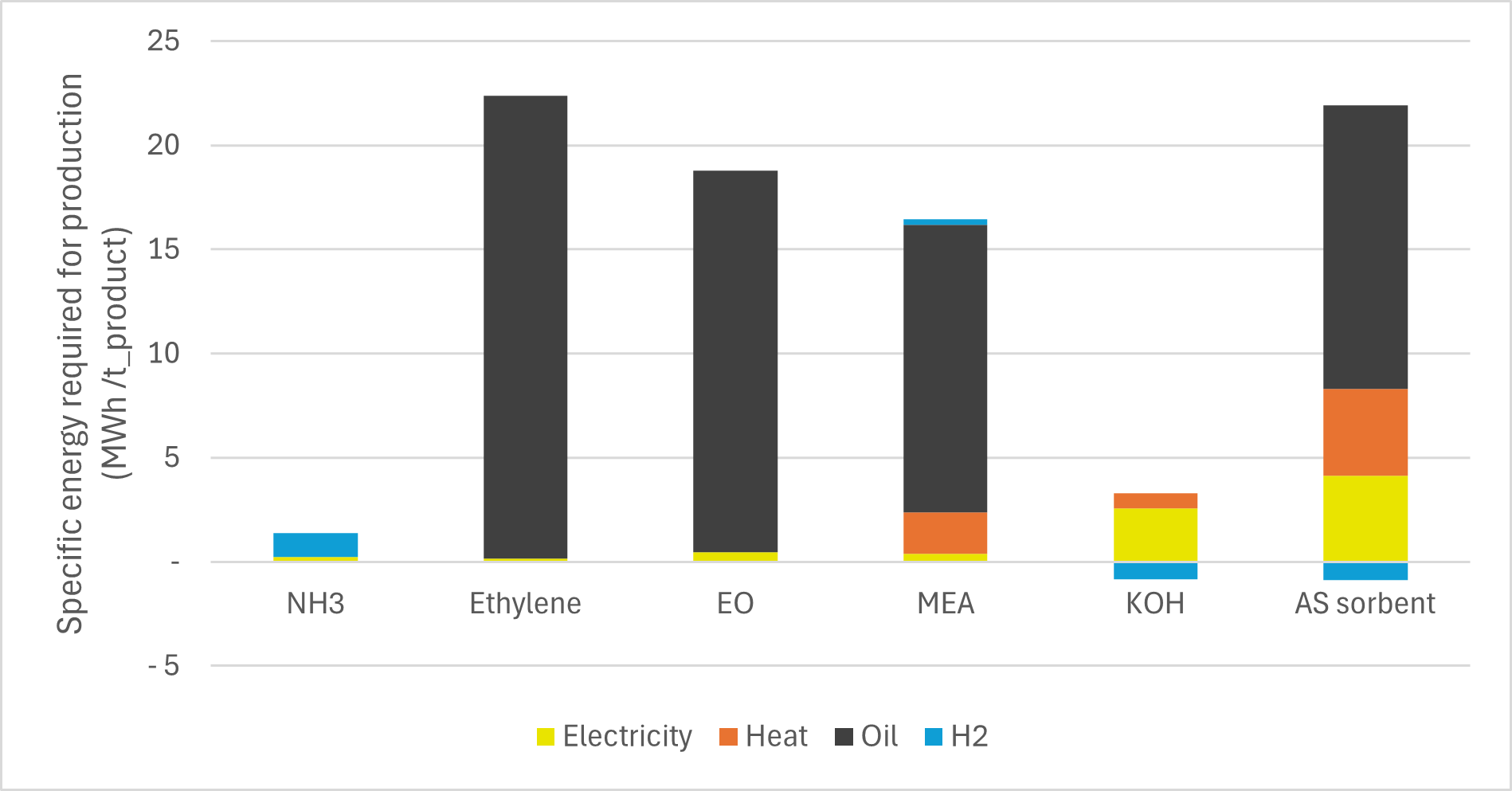}
    \caption{Specific energy requirement for production. This accounts for the whole production chain, including energy to produce each feedstock. The oil source (fossil, biofuel or efuel) is determined endogenously by the model. NH$_3$ = ammonia, EO = Ethylene oxide, MEA = Monoethanolamine, KOH = Potassium hydroxide, AS sorbent = Amine (PEI) -silica solid sorbent.}
    \label{fig:spec_eng}
\end{figure}

The specific energy inputs of solvent and sorbent production vary substantially (Figure \ref{fig:spec_eng}). MEA requires a large amount of oil due to its synthesis requiring ethylene, which is a petroleum derivative mainly produced from steam cracking of naphtha or ethane today \cite{POSCH201123}. The oil source differs substantially in the scenarios: it is fossil if a compensation by negative emissions elsewhere is possible, otherwise it is dominated by efuels if biomass is limited, or by biofuels if not. MEA is further used in the amine-silica sorbent synthesis, transmitting this high oil consumption. The Chlor-alkali process involved in producing KOH requires electricity input, but the overall energy requirement remains substantially lower than for the two other solvents, and hydrogen emerges as a by-product.

%In this process, 80\% of this oil is used as a feedstock, though, while the remaining 20\% is burnt to fuel the process \cite{REN_steam_cracking}.

%Figures \ref{fig:LCOC} and \ref{fig:Energy} showcase the effects of solvents and sorbents production on the LCOC and the energy requirement to capture carbon. Calculations of LCOC baselines as well as assumptions for energy for CC operation are presented explicitly in the Supplementary information. Cost calculations are based on formulas presented in Section \nameref{cost_assessment}.

% \subsection*{Effect on carbon capture costs and energy use}

\begin{figure}[htp]
    \centering
    \includegraphics[width=0.7\linewidth]{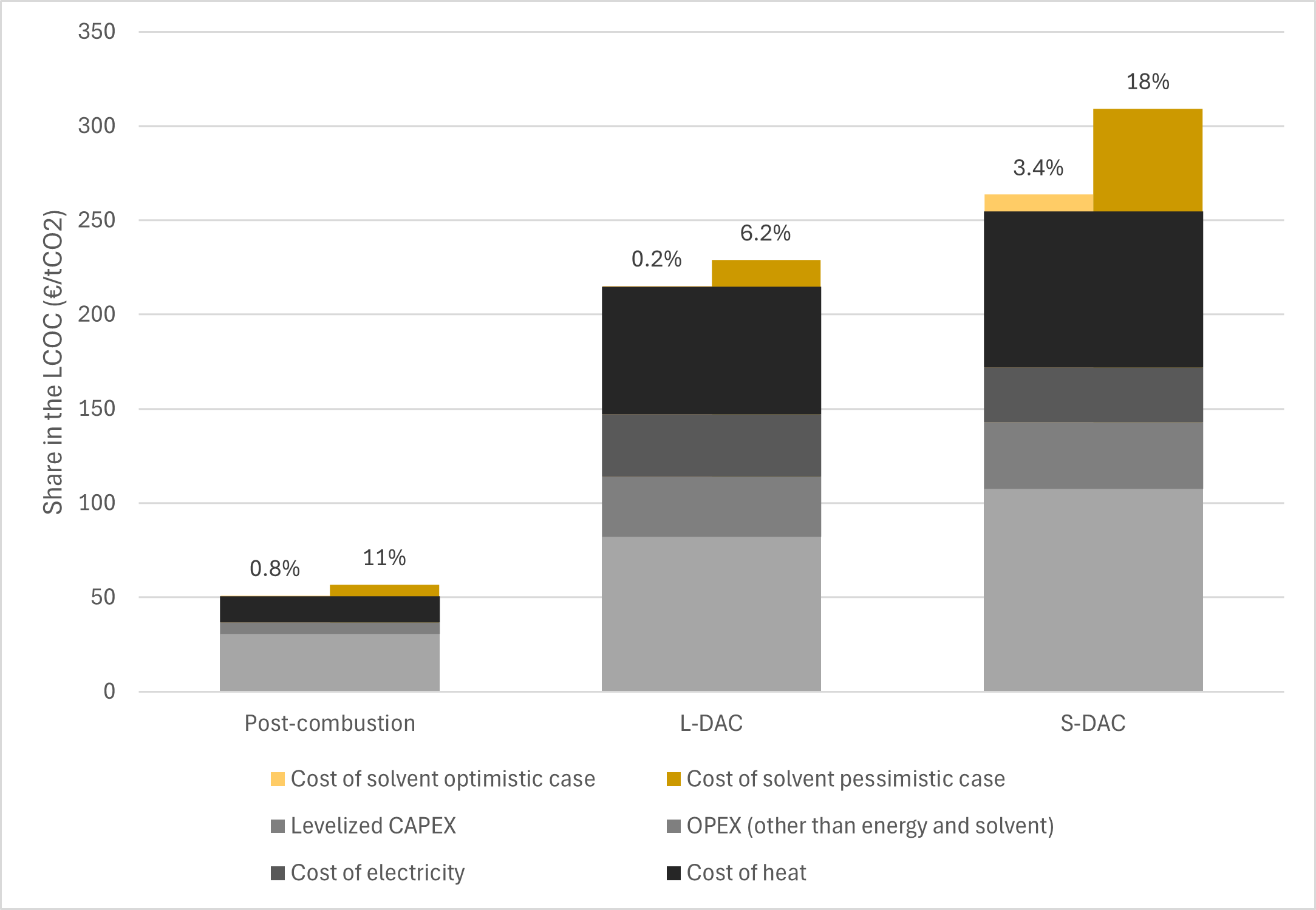}
    \caption{LCOC breakdown for each carbon capture technology. For post-combustion, only the carbon capture infrastructure cost is included, as carbon is considered a by-product. Since there are several post-combustion processes in the model, the baseline LCOC for this technology, without solvents, is determined by averaging the LCOCs calculated for each of them.}
    \label{fig:LCOC}
\end{figure}

\begin{figure}[htp]
    \centering
    \includegraphics[width=0.7\linewidth]{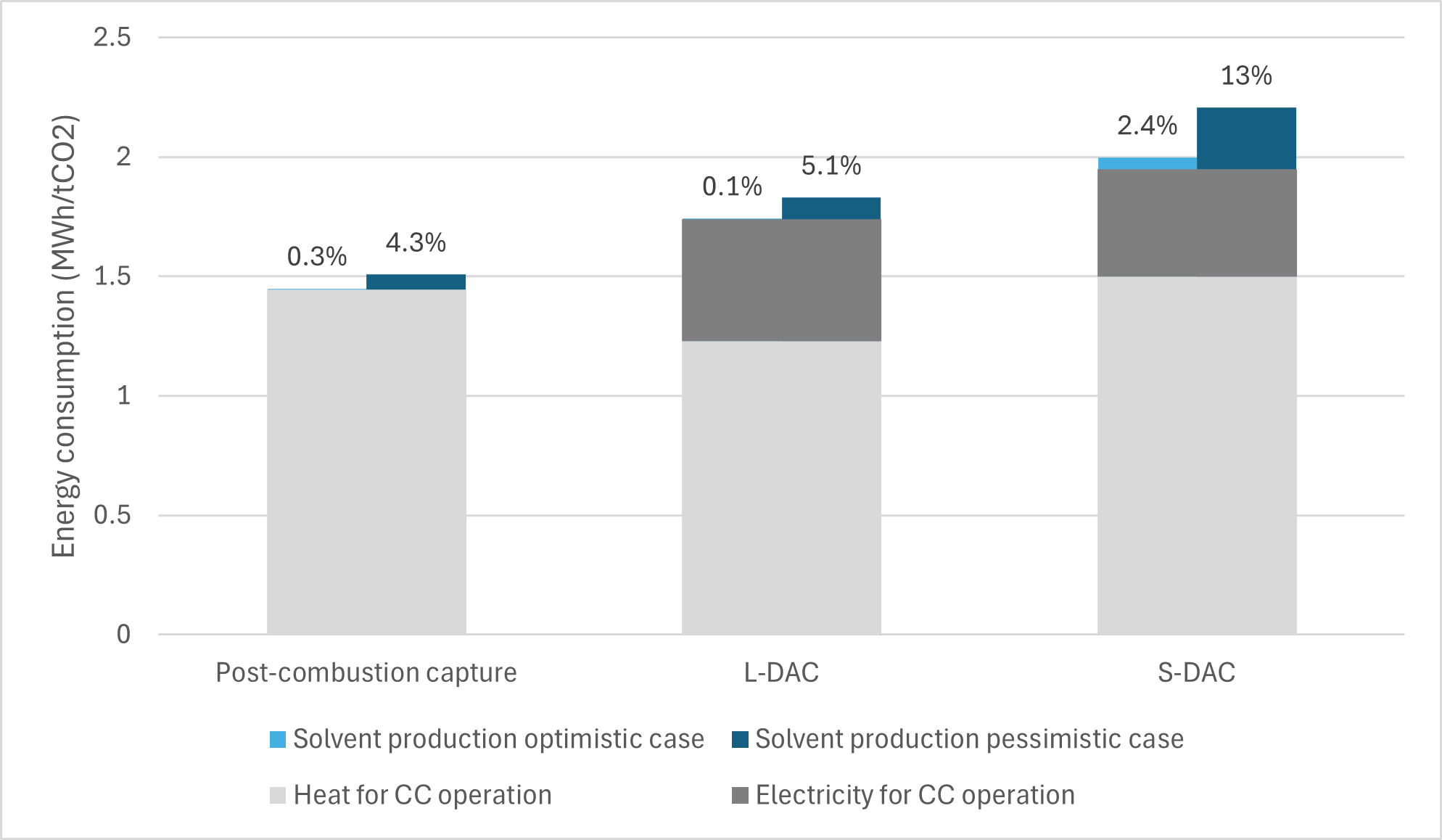}
    \caption{Energy consumption breakdown for each carbon capture technology}
    \label{fig:Energy}
\end{figure}

The effect of accounting for solvent production uncertainty on LCOC (Fig.~\ref{fig:LCOC}) and energy consumption (Fig.~\ref{fig:Energy}) differs between the three technologies, which affects their relative cost-competitiveness. While solvents contribute only marginally to the cost of post-combustion and L-DAC, by 6~\texteuro/tCO$_2$ (11\%) and 14~\texteuro/tCO$_2$ (6\%), respectively, S-DAC sorbents contribute to up to 55~\texteuro/tCO$_2$ (18\%) of LCOC and up to 0.3~MWh/tCO$_2$ (13\%) of energy consumption. Thereby, the cost advantage of L-DAC is further enhanced. If solvent and sorbent consumption falls in the optimistic range, effects on LCOC and energy use are small to negligible for all carbon capture options.

\begin{figure}[htp]
    \centering
    \includegraphics[width=0.8\linewidth, height=0.6\linewidth]{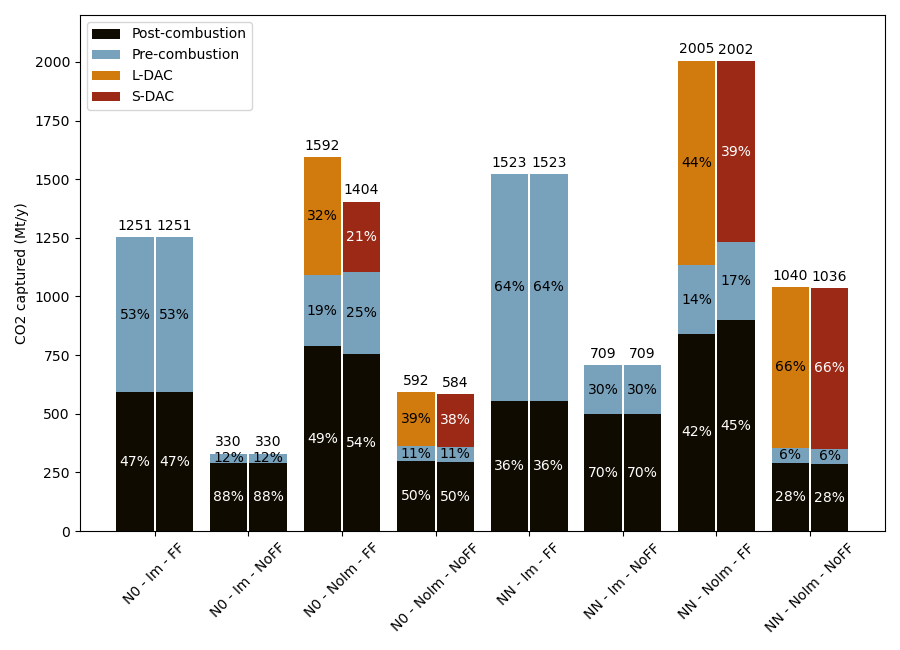}
    \caption{Amount of CO$_2$ captured by each type of technology. For each scenario, the bars to the left represent the case where S-DAC is excluded from the system, and the bars to the right, where L-DAC is excluded. As the capacities of these processes barely change between the base case and the case with solvent production, only the base case status is presented here. The labels represent the different scenarios: N0 = Net zero emission target, NN = Net negative, Im = with import of biomass, NoIm = No import of biomass, FF = Fossil fuels included, NoFF = No fossil fuels.}
    \label{fig:cc_share_ldac}
\end{figure}

The overall system effects also depend on the magnitude of deployment of the different carbon capture options. In the scenarios, carbon capture deployment exhibits a large range between 330~MtCO$_2$ in a net-zero scenario where fossil fuels are completely phased out, and 2005~MtCO$_2$ in a net-negative scenario including fossil fuels. If biomass is readily available, post- and pre-combustion carbon capture dominate (Fig.~\ref{fig:cc_share_ldac}), with a negligible effect (0-0.63\%) and (0-0.37\%) of solvent production on system costs and energy use respectively.

If, however, biomass is limited to medium domestic residue potentials only, DAC is used to supply carbon for fuel production (CCU) in the non-fossil scenario, or for negative emissions (CCS) if fossils are allowed. In scenarios where fossil fuels are allowed, more carbon capture cost-effectively compensates for fossil emissions, and thus, because more solvents are used, the effect of solvents is larger than in the non-fossil scenarios.

\begin{figure}[htp]
    \centering
    \includegraphics[width=0.8\linewidth]{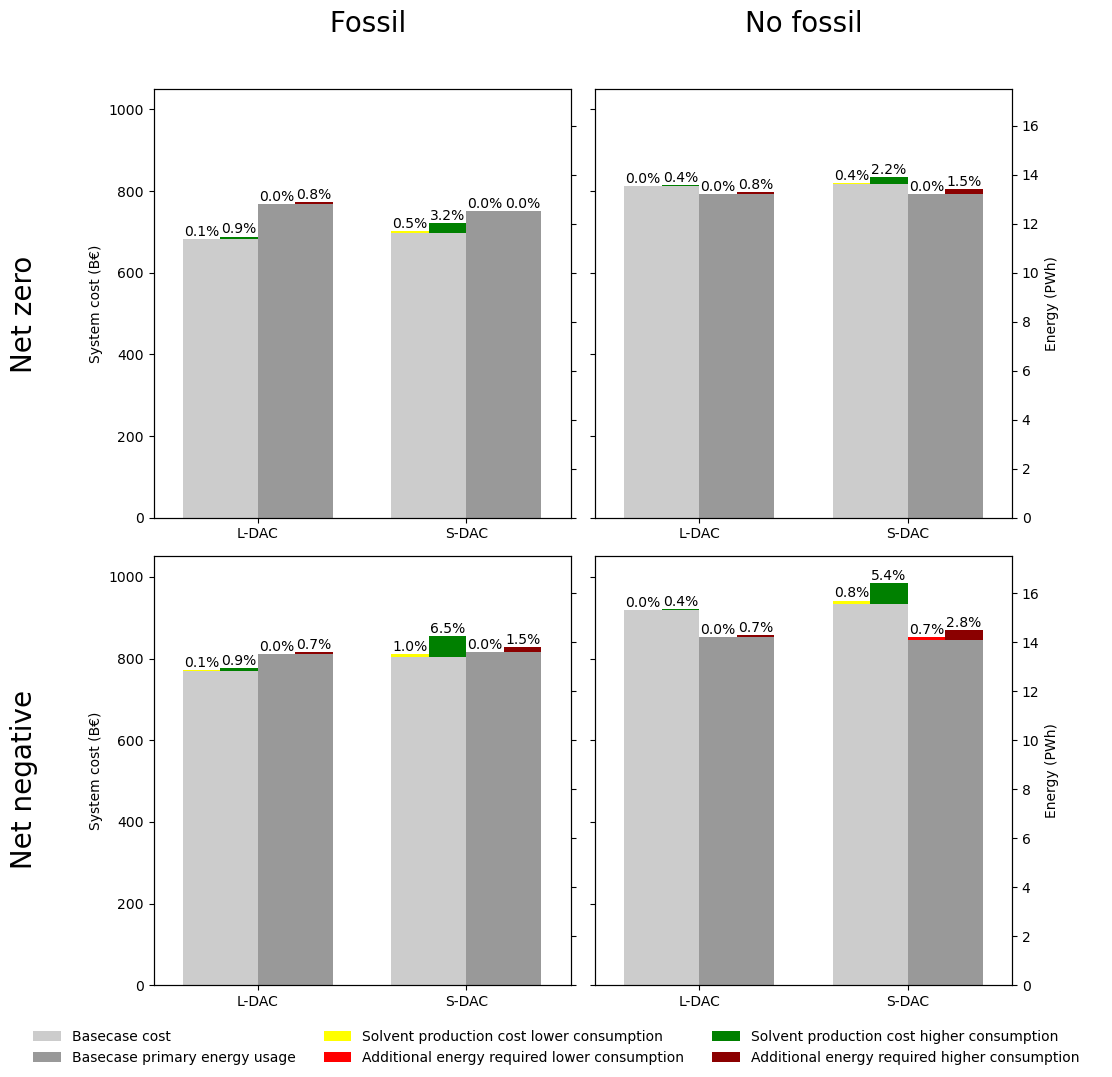}
    \caption{Impacts on the energy system of implementing solvent and sorbent production for carbon capture compared to a system in which this is not implemented. Scenarios with limited biomass availability are not showcased here.}
    \label{fig:Fossil and Emission target}
\end{figure}

Of the two DAC options, L-DAC is more cost-competitive, in which case solvent costs contribute to at most 0.9\% of the total energy system cost and to 0.8\% of energy demand (Fig.~\ref{fig:Fossil and Emission target}). If S-DAC is provided as the only DAC option (as is common in many ESM studies \cite{Millinger, PICKERING20221253, LUT}), the system is more clearly affected by sorbent uncertainty, with a total system cost contribution from sorbent production alone of 5.4-6.5\% (up to 52~billion~\texteuro) and an energy demand contribution from sorbent production of 1.5-2.8\% in the net-negative scenarios.

The main reason for the large uncertainty for S-DAC sorbent cost is the uncertain sorbent replacement rate. While L-DAC solvents have been tested on an industrial scale in the Kraft pulping process which narrows down the uncertainty also for L-DAC usage \cite{madhu2021understanding, KraftProcess, Sanz-perez, KEITH}, S-DAC sorbent degradation is still poorly understood \cite{deutz}.

DAC is a novel technology with a technology readiness level of 6-7, and thus the processes have yet to be tested on a large scale. Moreover, this study assumes a PEI-silica sorbent but other proprietary sorbents under development might perform better. Results in this study indicate that there is a strong cost incentive to develop sorbents that efficiently capture carbon and are resistant to degradation, but there is also a trade-off between sorbent reactivity, robustness and production cost \cite{Zhu2022}.

\begin{figure}[htp]
    \centering
    \includegraphics[width=0.8\linewidth]{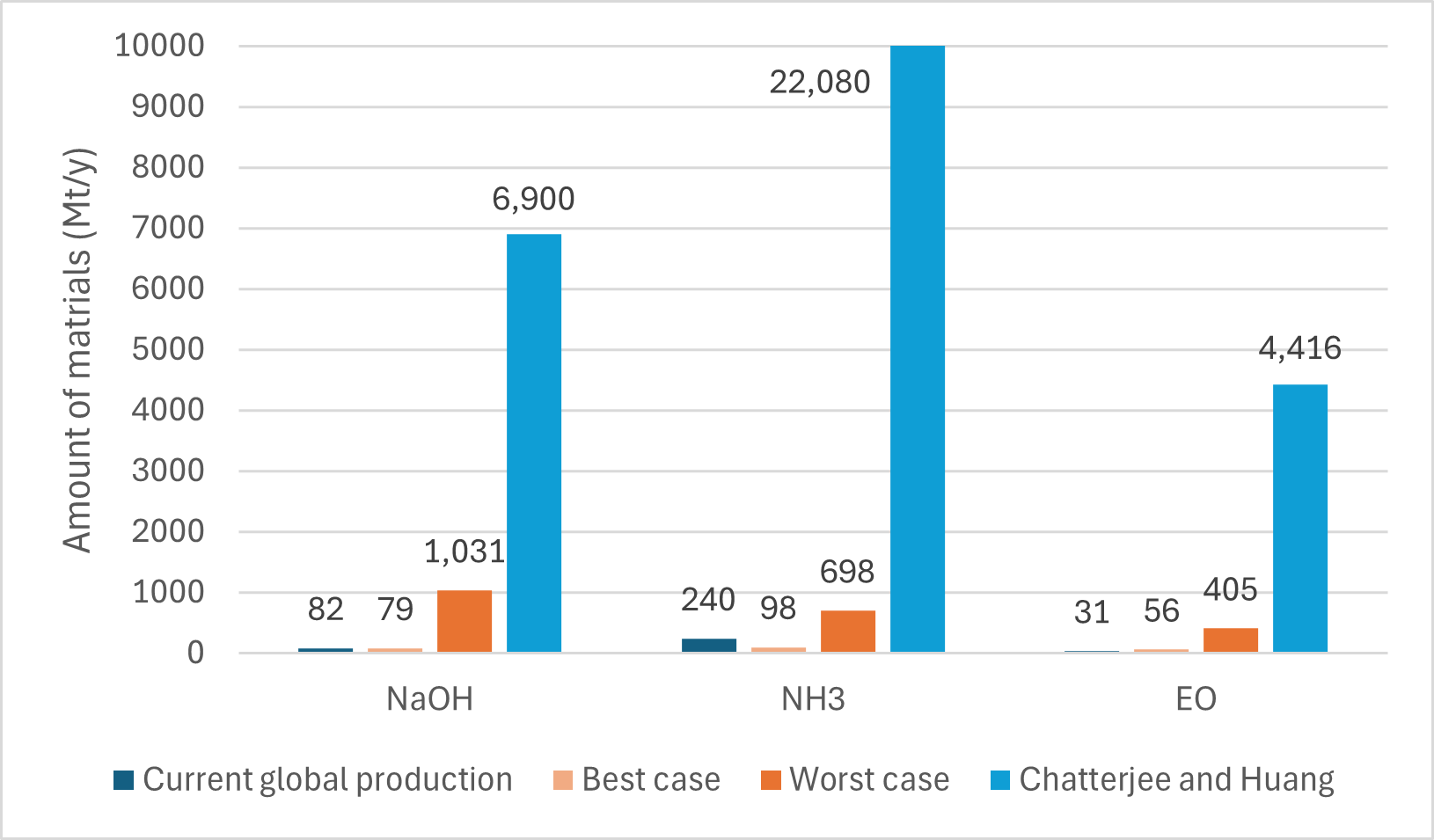}
    \caption{Material consumption with this study's assumptions for a DAC scale-up to 30 GtCO$_2$/y}
    \label{fig:material_cons}
\end{figure}

Although solvent production for carbon capture may require a substantial scale-up of production capacities (Tab.~\ref{tab:Material consumption}), the high material demand estimated by Chatterjee \& Huang \cite{Unrealistic} could not be confirmed (Fig.~\ref{fig:material_cons}), due to much lower solvent and sorbent consumption and much more efficient MEA production even in the pessimistic case. While ethylene, ethylene oxide and ammonia are already produced in much higher quantities than is demanded in the scenarios, the production of MEA and KOH may need to scale up by more than one order of magnitude, should global demand develop similarly to the obtained European demands. This is however little compared to the capacity expansion of other technologies such as solar, wind power or electrolysis in the assessed scenarios. For PEI, on the other hand, no large-scale production capacity exists today, which could present a bottleneck and a supply uncertainty in the expansion phase of S-DAC. Polymeric materials similar to PEI are however produced in large quantities globally, suggesting that such development is feasible. The scale-up of KOH and NaOH should present little challenges \cite{madhu2021understanding}, as they are derived from KCl and NaCl salts that are very abundant on Earth, and rely on the Chlor-Alkali process (electrolysis of brine), which is a common industrial process.

%to be needed by Realmonte and co-workers to model results shows a much smaller issue than expected. This difference can be explained by a combination of factors, but mainly a much lower solvent consumption assumption for both DAC technologies, from 170-290 t/t$_{CO_2}$ to 0.4-38 and 2.3-14 for L- and S-DAC respectively in this study, and a lower material and energy consumption for MEA production, especially of ammonia, from 3.2 to 0.25 t/t$_{MEA}$.

% ethylene could benefit from more renewable pathways being pursued instead of the conventional steam cracking. 
% Scaling the amount of materials that would be required for a 30 Gt carbon capture capacity is shown in Figure \ref{fig:material_cons}.

\begin{table}[htp]
\centering
% \begin{footnotesize}
\caption{Material requirement for the lower and the higher solvent consumption. The values are the highest material usage of all 8 scenarios and of either the S-DAC or L-DAC system.}
% \resizebox{\textwidth}{!}{
\begin{tabular}{lcccccc} \hline \rowcolor[HTML]{EFEFEF}
\begin{tabular}[l]{@{}l@{}} Material req. [Mt/y]\end{tabular} & 
MEA & KOH & PEI & EO & Ethylene & NH$_3$\\ \hline
\begin{tabular}[l]{@{}l@{}} Optimistic case \end{tabular}& 
2 & 2.1 & 0.9 & 1.5 & 1.3 & 2.6  \\
\begin{tabular}[l]{@{}l@{}} Pessimistic case \end{tabular}& 
14.4 & 27.5 & 5.4 & 10.8 & 8.9 & 18.6 \\
\begin{tabular}[l]{@{}l@{}} Global production \end{tabular} & 
2.5$^1$ \cite{deutz} & 9$^2$ \cite{oxford_institute}& 0.01 \cite{deutz} & 31 \cite{statista_ethylene_oxide_2023} & 225 \cite{statista__ethylene_2023} & 240 \cite{globaldata_ammonia} \\ \hline
\end{tabular}%}
\vspace{1ex}
% {\raggedright \\
\\ 1: Global production of all three EAs. \\
2: For NaOH, the global production is of 82 Mt/y \cite{eu_bat_ch_al}
 % \par}
\label{tab:Material consumption}
% \end{footnotesize}
\end{table}

% \begin{itemize}[noitemsep]
%     \item comparison to other studies
%     \item reflection on uncertainties
%     \item factors other than solvents that may affect competitiveness
%     \item trade-off high temp heat and solvent
%     \item implications for modelling
%     \item limitations
% \end{itemize}

DAC costs are subject to large uncertainties in literature resulting in cost projections for different DAC options often overlapping \cite{young2023cost}. Some studies have projected lower costs for S-DAC than for L-DAC mainly due to the lower temperature heat demand allowing for waste heat usage \cite{Fasihi}, but other studies find sorbent cost to drive the opposite outcome \cite{OZKAN2022103990}. 

While S-DAC has a larger disadvantage compared to L-DAC regarding both cost and energy use when it comes to sorbent and solvent production, L-DAC has other challenges. In particular, it requires about 900$^o$C heat input \cite{KEITH} for solvent regeneration compared to about 100$^o$C for S-DAC \cite{Beuttler_Charles_Wurzbacher_2019}. L-DAC therefore needs a heat source utilising methane (as assumed in this study), hydrogen or possibly electricity \cite{McQueen2021}, while S-DAC can use less costly options such as waste heat from industry (to the extent that this limited resource is available) \cite{young2023cost}, geothermal heat as Climeworks Iceland plants Orca and Mammoth do \cite{Climeworks_orca,Climeworks_mammoth}, or industrial heat pumps \cite{Breyer2020, Terlouw2024}. Moreover, from a life-cycle perspective, L-DAC has been found to have a higher environmental footprint than S-DAC if natural gas is assumed for heat supply and due to the displacement of substantial mass flows in the process \cite{madhu2021understanding}. L-DAC also consumes significant amounts of fresh water due to evaporation of the diluted solvent in contact with air and require up to 6 times more land area than S-DAC \cite{bergman2021case, madhu2021understanding} to implement its bulky infrastructures. Conversely, S-DAC uses dry sorbents and its modular infrastructure makes it less area intensive \cite{madhu2021understanding}. 

Energy system studies commonly have included S-DAC as the sole DAC option \cite{LUT,PICKERING20221253,Millinger}, with the advantage that the heat source can be supplied by low cost waste heat or through heat pumps \cite{Fasihi}. This study finds that S-DAC sorbent induced energy system cost uncertainties are substantial, and thus presents a trade-off with the more flexible heat source which has been mentioned as a strong argument for S-DAC.

\subsection{Limitations}
The choice of solvents and sorbents studied in this model was dependent on available data. For sorbents especially, there are multiple options being considered for use, and price forecasts exhibit large variations \cite{young2023cost, SIEVERT2024979, NASEM, WANG2012319}, with spans exceeding the costs assumed in this study, which may therefore even be underestimated.

Solvents associated with pre-combustion were excluded from the study as reports suggest that the degradation of solvents used in this technology are minimal \cite{DUVALDACHARY2023113415}. Alternative CCS technologies such as oxy-fuel combustion, chemical looping and membrane options were not considered as they generally avoid the use of solvents or sorbents and exhibit lower technology readiness levels.

%\textbf{Post-combustion lower cost if CO2 by-product, but biomass limited resource and carbon may become the main product \cite{Millinger}. , etc.}

\section{Conclusions}
Although carbon capture costs remain dominated by capital and operational energy expenditures, solvent cost may be significant, especially for DAC using solid sorbents. The cost and energy use uncertainty of solvents for post-combustion carbon capture as well as L-DAC and S-DAC was assessed and was found to affect their internal cost-competitiveness and potentially increase energy system cost significantly. For post-combustion carbon capture and L-DAC, solvent uncertainties present negligible challenges on the energy systems level. In contrast, S-DAC solvent uncertainties were found to represent a share in LCOC of up to 18\%, and contribute to total energy systems cost by as much as 6.5\% (52~billion~\texteuro) in net-negative (-110\%) scenarios. Solid sorbent degradation and replacement accounts for large cost uncertainties, providing strong incentives to develop robust sorbents which are still reactive and affordable. Scale-up of solvent and sorbent production was estimated to be uncritical compared to the observed scale-up of many other technologies such as VRE and electrolysers.

S-DAC is modular and therefore is more flexible in terms of unit size, requires lower temperature heat to regenerate the sorbent, and it has potentially lower environmental impact, which offers advantages compared to L-DAC. There is thus a trade-off between these advantages and the additional cost uncertainty regarding sorbents. 

Energy system modelling studies have thus far not included details on sorbents and solvents but large amounts of carbon capture are obtained in climate mitigation scenarios. While this was found to be uncritical for post-combustion carbon capture, uncertainties especially for S-DAC are substantial and need to be considered in modelling assessments. This is an issue especially in climate mitigation scenarios with low biomass availability and where S-DAC plays a large role, in which case sorbent-induced cost uncertainties are the largest.

\section*{Supplementary information}
\label{suppinfo}
Supplementary information consisting of a spreadsheet that contains the entire data collection carried out in this work as well as some of the main calculations about the impacts on the technologies can be found at this web address: %\url{https://zenodo.org/records/13312324?token=eyJhbGciOiJIUzUxMiJ9.eyJpZCI6ImZjMDIyY2Q1LThiMTEtNGI2ZC04ZDNhLTYwZWJmNjU3NGM3YyIsImRhdGEiOnt9LCJyYW5kb20iOiJhNzY5ZTFkOWRlNjJiNTZlNjQzNTE5MjE5ZTM0YmJjNCJ9.fd0ruh6x rBIkCpDcZvr8fo\_8gARrbFJ-RkcqXYSDttEp\_tBjoqIoR6NqAgq3P-LhwLKixJZP9iX d7TGruPwElw}
\url{https://doi.org/10.5281/zenodo.13312323}

\section*{Code availability}
\noindent
The code used can be found at \url{https://github.com/humpage/pypsa-eur-sec/tree/solvent} \\
Technology data used can be found at \url{https://github.com/humpage/technology-data/tree/biopower}

\section*{Acknowledgements}
\noindent
We thank Fredrik Hedenus for valuable input to this work.

We acknowledge funding from the Swedish Energy Agency, project number 2021-00067. The computations were enabled by resources provided by Chalmers e-Commons at Chalmers.

\bibliographystyle{IEEEtran}

\bibliography{citations}

\end{document}